\documentclass[12pt]{iopart}

\newcommand{\ai}{\'{\i}}               %% \ai{} to accent the i
\renewcommand {\slash}[1]{#1 \!\!\! /} %% \slash{} for "A slash"

\usepackage{epsfig}  %% To include eps figures 
\usepackage{comment} %% For large regions of commented text
\usepackage{amssymb} %% For the AMS symbols

\def\lambdabar{\protect\@lambdabar}
\def\@lambdabar{%
\relax
\bgroup
\def\@tempa{\hbox{\raise.73\ht0
\hbox to0pt{\kern.25\wd0\vrule width.5\wd0
height.1pt depth.1pt\hss}\box0}}%
\mathchoice{\setbox0\hbox{$\displaystyle\lambda$}\@tempa}%
{\setbox0\hbox{$\textstyle\lambda$}\@tempa}%
{\setbox0\hbox{$\scriptstyle\lambda$}\@tempa}%
{\setbox0\hbox{$\scriptscriptstyle\lambda$}\@tempa}%
\egroup
}

\begin{document}  

\title[Quantum vs classical scattering of 
       Dirac particles]
      {Quantum versus classical scattering of 
       Dirac particles by a solenoidal magnetic field
       and the correspondence principle}    

\author{Gabriela Murgu\ai{}a,
        Mat\ai{}as Moreno
        and Manuel Torres
       }  

\address{Instituto de F\ai{}sica, UNAM.
         Apartado postal 20364, 01000, M\'exico, D.F. Mexico.}  

\date{\today}  

\eads{\mailto{murguia@fisica.unam.mx},
      \mailto{matias@fisica.unam.mx},
      \mailto{torres@fisica.unam.mx}
     }

%%% Abstract %%%   
\begin{abstract}  
    
We present a detailed analysis of the scattering of charged particles
by the magnetic field of a long solenoid of constant magnetic flux and
finite radius. 
We study the relativistic and non-relativistic quantum and classical
scenarios.
The classical limit of the perturbative quantum expressions,
understood as the Planck's limit (making $\hbar$ going to zero) is
analyzed and compared with the classical result.  The classical cross
section shows a general non-symmetric behavior with respect to the
scattering angle in contradistinction to the quantum calculations
performed so far.
The various regimes analyzed show that the quantum cross sections do
not satisfy the correspondence principle: they do not reduce to the
classical result in any considered limit, an argument in favor of the
interpretation of the process as a purely quantum phenomenon.
We conclude that in order to restore the classical correspondence of
the phenomenon, a complete non-perturbative quantum calculation for a
finite solenoid radius is required.

\end{abstract}  


\maketitle
  
%%% Section: Introduction %%%   
\section{Introduction}   
\label{sec:Introduction}  

The interactions of charged particles with magnetic fields have been
widely studied. Important developments, both theoretical and
experimental have appeared continuously, as those in Solid State
Physics~\cite{Lee_1989}, and Particle Physics and
Cosmology~\cite{Pal_2002, Grasso_2000, Wibig_2004}.
One of the most important effects with magnetic fields is perhaps the
Aharonov-Bohm (AB) effect~\cite{AB}. Since it was proposed, several
experiments and studies about it have been developed~\cite{Chambers,
Osakabe}.
A lot of work has been done about this particular
effect~\cite{Peshkin-Tonomura, Nambu, Bagrov:2000hv, Bagrov:2002sc},
as well as in those that concern to the general properties of magnetic
fields and its interaction with matter~\cite{Gavrilov} and a few of
them deal with the classical aspect of the
AB-effect~\cite{Ley-Koo_2001}.
It is frequently mentioned that the AB-effect is a pure quantum effect
and that it represents one of the hardest tests that quantum
mechanics has successfully approved.
In the zero radius limit of the solenoid the effect is purely quantum,
because in $\hbar \rightarrow 0$ limit, the AB differential cross
section (DCS) cancels exactly.
However, one may wonder if there is some relation between the
classical and quantum regimes in the case of finite solenoid radius.
We pose to question if there is any particular limit in which the
quantum result reduces to the classical one, otherwise we have an
example in which the correspondence principle fails in its $\hbar
\rightarrow 0$ version.
In this paper we deal with one of this particular problems, the
elastic scattering of charged spin-1/2 particles by an external
solenoidal magnetic field.
We carefully analyze both, the classical and quantum relativistic and
non-relativistic scenarios.
For the field theoretic calculation we argue that renormalization
effects do not modify our conclusion.

The classical correspondence of the quantum scattering of particles by
magnetic fields seems to be a puzzle still to be solved. Is this
another pure quantum phenomena as was suggested before in Landau and
Lifshitz book~\cite{LL} in the zero-radius limit?  
Since the quantum field theory success in the Coulomb scattering, in
which the Rutherford classical cross section is recovered, the
following question arises: Can the classical limit of the scattering
by a solenoid be taken for granted?  AB-effect is a clear example that
we should be prepared for surprises.

It is well known that the Planck's limit, $\hbar \rightarrow 0$, is
delicate~\cite{Balian-Bloch, Berry_1972, book:Gutzwiller}. The particular
scattering problem analyzed in this paper is an example of this
because the classical expressions are not obtained from the quantum
ones.
The search of a general prescription to obtain classical limits is
still an open problem.

The paper is organized as follows.  
In section~\ref{sec:Classical_Result} we calculate the classical DCS
of the scattering problem, although direct, this result is not readily
found in the literature.  There, we discuss two important limiting
cases, which correspond to a uniform and constant magnetic field, and
a thin solenoid of fixed flux (AB-case).  In
section~\ref{sec:Quantum_Result} we present the calculations in both,
relativistic~\cite{Murguia_2003} and non-relativistic~\cite{AB, LL}
quantum regimes. There we stress the consistently and completely
symmetric behavior of the quantum DCS in the scattering angle.  In
section~\ref{sec:Analysis} we compare the classical and quantum
results and we extend the discussion of the effect produced in the
$\hbar$ dependence of the relativistic perturbative result to all
orders in $\beta = e\Phi/2\pi c$.  Finally, in
section~\ref{sec:Conclusions} we present our conclusions.

%%% Section: Classical result %%%  
\section{Classical Cross Section}  
\label{sec:Classical_Result}  

The classical problem of the scattering of a charged particle by the
magnetic field of a long solenoid of radius $R$ and constant magnetic
flux is analyzed first.
Besides we will show in section~\ref{sec:Quantum_Result} that our
results are relativistically correct and due to the delicate nature of
the classical-quantum comparison, it is of basic importance to have a
correct result for the classical differential cross section of charged
particles by magnetic fields.

%% Subsection: The scattering angle 
\subsection{The scattering angle}

\begin{center}
\begin{figure}[ht]
\epsfig{file=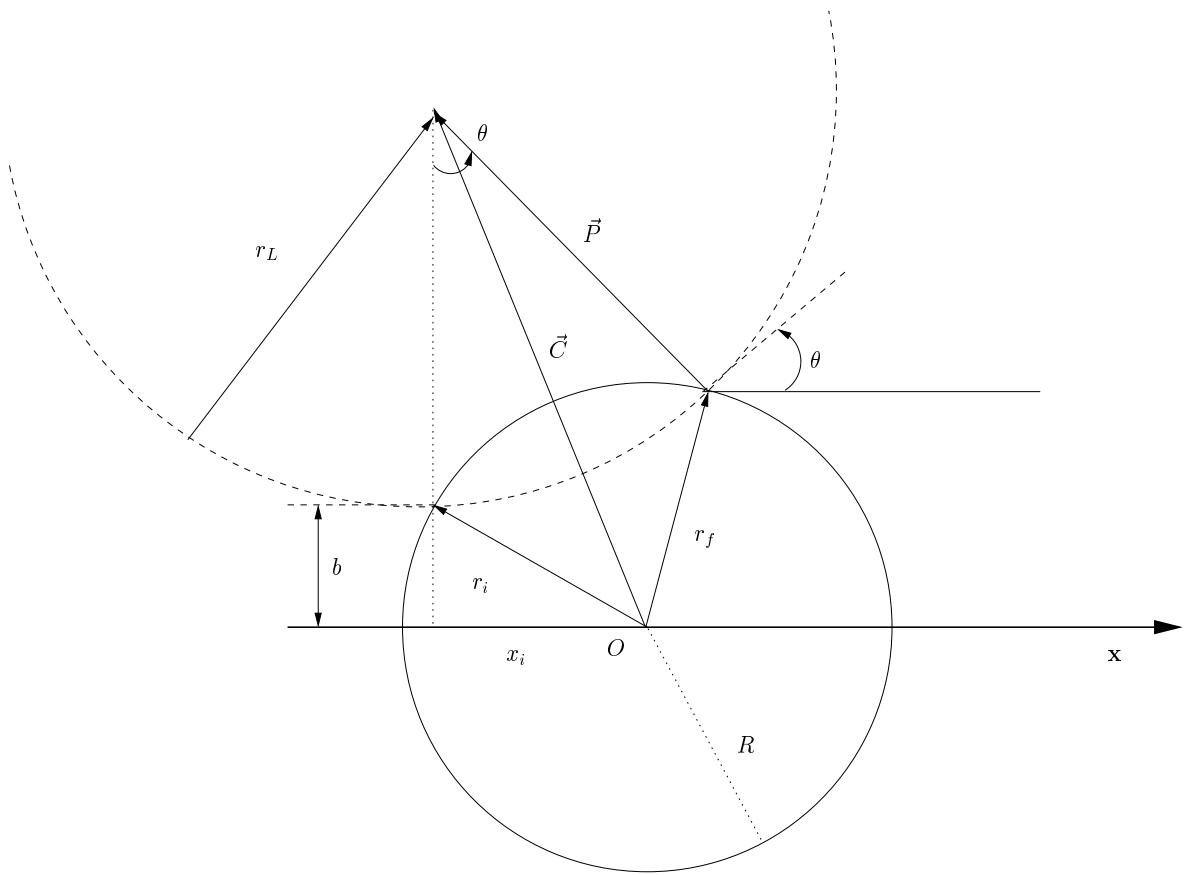,bb=127 158 821 668,angle=0,width=\linewidth,clip=}
 \caption[Classical scattering by a solenoidal magnetic field.]
         {Classical scattering of a charged particle by the magnetic 
           field of a solenoid.}
 \label{figure:dispersion_clasica_1}
\end{figure}
\end{center}

In the figure~\ref{figure:dispersion_clasica_1}, the scattering of a
negatively charged particle coming from the left of the solenoid with
velocity along the $x$-axis is schematically shown. This is a general
situation due to the axial symmetry of the problem.  The uniform and
constant magnetic field $B$ differs from zero only inside of the
solenoid and is defined to point out-wards of the paper plane. The
charged particle enters the magnetic field region with an impact
parameter $b \in [-R,R]$, and the points in which it enters and leaves
the solenoid are denoted by the position vectors $\bi{r_i}$ and
$\bi{r_f}$ respectively. It is assumed that the particle does not
radiate while it is interacting with the magnetic field, so its
trajectory inside the solenoid will be an arc of a circumference of
radius $r_L$ centered at $\bi{C}$ respect to the axis of the solenoid.
The particle leaves the solenoid at point $\bi{r_f}$ corresponding to
a scattering angle $\theta$ measured respect to the horizontal
(incident direction). The non radiation assumption implies that
$\theta \in [0, 2\pi)$.

%%%%%%%%%%%%%%%
%%%%%%%%%%%%%%%

With the geometry shown before, 
$$
\bi{r_i} = (-\sqrt{R^2-b^2},b).
$$ 
$\bi{r_f}$ belongs to both
circumferences, the one that describes the solenoid of radius $R$, and
the one of Larmor radius $r_L$ that describes the trajectory of the particle
inside the solenoid. Then
$$
\bi{r_f} = \left(\frac{\sqrt{R^2 -b^2}\ ({r_L}^2 - R^2)}{R^2 + 2br_L + {r_L}^2},
                   b - \frac{2(b-R)(b+R)r_L}{R^2 + 2br_L + {r_L}^2}
            \right).
$$
The vector $\bi{P} = \bi{r_f}-\bi{C}$,
gives us information about the scattering angle $\theta$ as a function
of the impact parameter $b$ accordingly to
\begin{eqnarray}
\theta(b) & = & \arctan{(|P_x|/|P_y|)} \nonumber \\
          & = & 2 \arctan{ \left( \frac{ \sqrt{R^2-b^2} }{b+r_L} \right) },
\label{eq:theta(b)}
\end{eqnarray}
because 
$$
\bi{C} = (-\sqrt{R^2-b^2},b+r_L).
$$ 
In fact there are two solutions for $\theta(b)$ corresponding to the
values of the two possible curvature concavities of the trajectories
of the particles depending on the sign of $eB$.  We determine the
sign of $eB$ in such a way that there is only one physical solution
for $\theta(b)$ which corresponds to that of
equation~(\ref{eq:theta(b)}).

Making use of the Newton's Second Law combined with the Lorentz's force,
$$
\frac{d\bi{p}}{dt} = -e \frac{\bi{p}}{mc} \times \bi{B},
$$
and using the fact that the momentum $\bi{p}$ of the particle is
at any given time perpendicular to the magnetic field $\bi{B}$, one can
show that
\begin{equation}
r_L = \frac{p c}{e B},
\label{eq:r_L}
\end{equation}
which is relativistically correct~\cite{Jackson}.  In these
expressions, $e$ and $m$ stand for the charge and mass of the
scattered particle, and as usual, $c$ denotes the speed of light.

%% Subsection: The differential cross section
\subsection{The differential cross section}

To obtain the differential cross section (DCS) of the classical scattering problem,
the impact parameter $b$ has to be expressed as a function of the
scattering angle $\theta$, {\it i.e.} equation~(\ref{eq:theta(b)}) has
to be inverted. Then, the general solution for the differential cross
section will be~\cite{book:Goldstein}
$$
\frac{d\sigma(\theta)}{d\theta} = \sum_i\left|{\frac{db_i(\theta)}{d\theta}}\right|,
$$
where $b_i(\theta)$ are the different 
branches of the solution.

Defining $\beta = e\Phi/2 \pi c $ and introducing the dimensionless parameters 
\begin{eqnarray*}
 \rho_b &=& b/R, \\
 \rho_L   &=& r_L/R = p R/2\beta,
\end{eqnarray*} 
that express the impact parameter and the Larmor radius in units of
the solenoid radius respectively ($\rho_b \in [-1,1]$, $\rho_L \in
[0,\infty)$), once the value of $eB$ has been fixed
and therefore the sign of equation~(\ref{eq:theta(b)}) determined,
the solutions for $\theta(\rho_b)$ are:
\begin{equation}
\rho_b^{\pm}(\theta,\rho_L) = -\rho_L \sin^2(\theta/2) \pm 
            \cos(\theta/2) \sqrt{ 1 - \rho_L^2\sin^2{(\theta/2)}}.
\label{eq:beta} 
\end{equation}

As indicated, the solutions $\rho_b^{\pm}(\theta,\rho_L)$
depend on the incident energy of the scattered particles. This is
taken into account by the parameter $\rho_L$, 
the relevant one in the 
classical DCS. It is related to the typical magnetic coupling $eB$ as
$$
\rho_L = \frac{1}{R} \, \frac{pc}{eB}.
$$ 
There are several physical limiting cases of the classical DCS with respect to
this parameter, corresponding to those shown in table~\ref{table:rho_limits}.
In all the cases $p$ has been taken as fixed with a finite value.
The particular case of constant magnetic flux $\Phi$ will be
considered here.
\begin{center}
\begin{table}
\caption{\label{table:rho_limits}Different limiting cases for the parameter $\rho_L$.}
\begin{indented}
\item[]\begin{tabular}{@{}ll}
\br
$\rho_L \rightarrow 0$ & ~ $R \rightarrow 0$ with $e,\Phi$ fixed \\
\hline
$\rho_L \rightarrow \infty$ & ~ $R \rightarrow \infty$ with $e, \Phi$ fixed \\
                            & ~ $R \rightarrow 0$ with $e, B$ fixed \\
                            & ~ $e \rightarrow 0$ with $R, B$ fixed ($\equiv\Phi$ fixed)\\
                            & ~ $B \rightarrow 0$ with $e, R$ fixed \\
\br
\end{tabular}
\end{indented}
\end{table}
\end{center}

For low energy, the Larmor radius is smaller than the solenoid one
and the trajectories of the particles can turn inside the solenoid. Therefore the
scattering angle $\theta$ will be found in the four quadrants: 
$\theta \in [0, 2\pi)$.
In this case, the relation
between $\theta$ and $b$ (equation~(\ref{eq:theta(b)})) is biunivocal,
hence the inverse has a unique physical solution given by $\rho_b^{+}$.
Thus the solution for $\rho_L < 1$ is
\begin{equation}
\rho_b(\theta,\rho_L < 1) = \rho_b^{+}(\theta,\rho_L) \mbox{ for } \theta \in [0, 2\pi).
\label{eq:beta_rho<1}
\end{equation}

On the other hand, for high energy the particles will be scattered in
the upper half plane only because their Larmor radius is larger than the
solenoid one.
Also notice that because $\rho_b$ is restricted to be real and to have
values between $-1$ and $1$, for this $\rho_L \geq 1$ case the square root
in equation~(\ref{eq:beta}) restricts the scattering up to a maximum
angle $\theta_{\mbox{max}} \leq \pi$ given by the trigonometric condition
\begin{equation}
\sin(\theta_{\mbox{max}}/2) = 1/\rho_L,
\label{eq:theta_max}
\end{equation}
and both solutions $\rho_b^{\pm}$ are physically acceptable.
In this case, the scattering angle will be found in the
first and second quadrants: $\theta \in [0, \theta_{\mbox{max}}]$.
The
$\theta - b$ relation is not biunivocal and both solutions
$\rho_b^{\pm}$ must be taken into account: 
\begin{equation}
\rho_b(\theta,\rho_L \geq 1) = \rho_b^{\pm}(\theta,\rho_L) \mbox{ for } \theta \in [0, \theta_{\mbox{max}}].
\label{eq:beta_rho>1}
\end{equation}

In both cases, $\rho_L < 1$ and $\rho_L \geq 1$, the expected
asymmetry in the scattering predicted by the Newton's Second Law for
the Lorentz's force will turn out.

Given $\rho_b(\theta,\rho_L)$,
it is straightforward to write down the classical DCS
for the scattering of a charged particle by the magnetic field
of a solenoid:
\begin{eqnarray}
\frac{1}{R} \frac{d\sigma_{\rho_L}(\theta)}{d\theta}
                       &=&   \left|{ \frac{\sin{\theta}}{2}
                                  \left({ \rho_L 
                                    + \frac{1 + \rho_L^2 \cos{\theta}}
                                           {2 \cos{(\theta/2)} \sqrt{ 1 - \rho_L^2\sin^2{(\theta/2)}}}
                                  }\right) 
                              }\right| + \nonumber \\
                        & &   \left|{ \frac{\sin{\theta}}{2}
                                  \left({ \rho_L 
                                    - \frac{1 + \rho_L^2 \cos{\theta}}
                                           {2 \cos{(\theta/2)} \sqrt{ 1 - \rho_L^2\sin^2{(\theta/2)}}}
                                  }\right) 
                              }\right| \Theta(|\rho_L| -1), 
\label{eq:secc_dif_clasica}
\end{eqnarray} 
where $\theta \in [0,2\pi)$ if $\rho_L < 1$ and
$\theta \in [0,\theta_{\mbox{max}}]$ if $\rho_L \geq 1$. 
$\Theta(x)$ is the usual Heaviside (unit step) function.

In general, this DCS presents an asymmetric behavior under the $\theta
\rightarrow 2\pi- \theta$ transformation.  According to
figure~\ref{figure:dispersion_clasica_1}, this behavior is related to
the reflection with respect to the $x-$axis.
Notice that when $e \rightarrow -e$ or $\rho_L
\rightarrow -\rho_L$ the same asymmetry results in
equation~(\ref{eq:secc_dif_clasica}), that is, all these
transformations are totally equivalent.

%% Subsection: Zero radius limit
\subsection{Zero radius limit (Low energy incident particles)}

As expected, the Lorentz's force produces in general a classical DCS
that is not symmetric with respect to the scattering angle $\theta =
0$.  Nevertheless, there is an interesting limit, $pR \rightarrow 0$
$(\rho_L \rightarrow 0)$, for which the cross section behaves
symmetrically with respect to the change $\theta \rightarrow 2\pi -
\theta$.  This case is equivalent to consider small energy incident
particles, hence backscattering is greatly favored:
$$
\left.{\frac{d\sigma}{d\theta}}\right|_{\rho_L \rightarrow 0} = \frac{R}{2}\left|{\sin(\theta/2)}\right|.
$$

\begin{center}
\begin{figure}[ht]
\epsfig{file=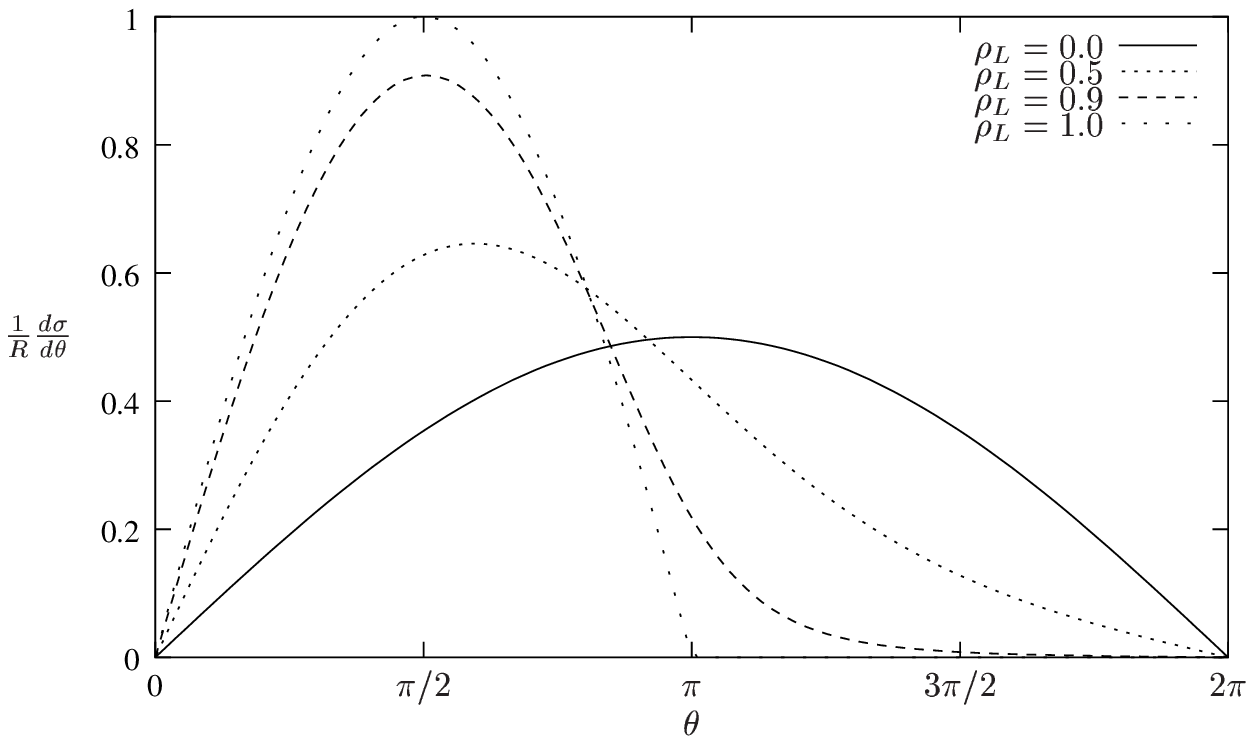,bb=119 453 478 667,angle=0,width=\linewidth,clip=}
 \caption[Classical cross section for the scattering by a solenoidal magnetic field for several values of
          $\rho_L = pR/2\beta \le 1$. ]
         {Classical cross section for the scattering by a solenoidal magnetic field for several values of
          $\rho_L = pR/2\beta \le 1$.}
 \label{figure:classical_CS_less}
\end{figure}
\end{center}

Figure~\ref{figure:classical_CS_less} shows how the DCS
behavior becomes symmetrical as the parameter $\rho_L$ approaches to zero,
whereas for values of $\rho_L$ below unity, the DCS results totally
asymmetric in the full range of $\theta \in [0, 2\pi)$.

%%%%%%%%%%%%%%%
%%%%%%%%%%%%%%%

%% Subsection: Big radius limit
\subsection{High energy incident particles}

Consider the limit $pR \rightarrow \infty$ $(\rho_L \gg 1)$ that, for
fixed magnetic flux (fixed $\beta$), corresponds to very high energy
incident particles.  In this case, the DCS as given by
equation~(\ref{eq:secc_dif_clasica}) diverges in the direction of the
maximum scattering angle $\theta_{\mbox{max}}$ (notice that the poles
of the cross section correspond to the zeros of $|d\theta/d\rho_b|$).
There is a manifestly asymmetric behavior of the cross section with
respect to the scattering angle when it changes from $\theta$ to $2\pi
- \theta$, and as long as the condition $\theta \leq
\theta_{\mbox{max}}$ is fulfilled, it is possible to show from
equation~(\ref{eq:secc_dif_clasica}) that the cross section reduces to
\begin{equation}
\left.{\frac{d\sigma}{d\theta}}\right|_{\rho_L \gg 1}  \approx
R \, \theta \frac{1+\rho_L^2}{\sqrt{4-\rho_L^2\theta^2}}
\mbox{ for } \theta \in [0, \theta_{\mbox{max}}]. 
\label{eq:secc_dif_clasica_small_rho}
\end{equation}

Notice that taking the limit $e \rightarrow 0$ in
equation~(\ref{eq:secc_dif_clasica}) is equivalent to consider $\rho_L
\rightarrow \infty$. And even more, for fixed solenoid radius $R$ this
result is also obtained when $eB \rightarrow 0$.

\begin{center}
\begin{figure}[ht]
\epsfig{file=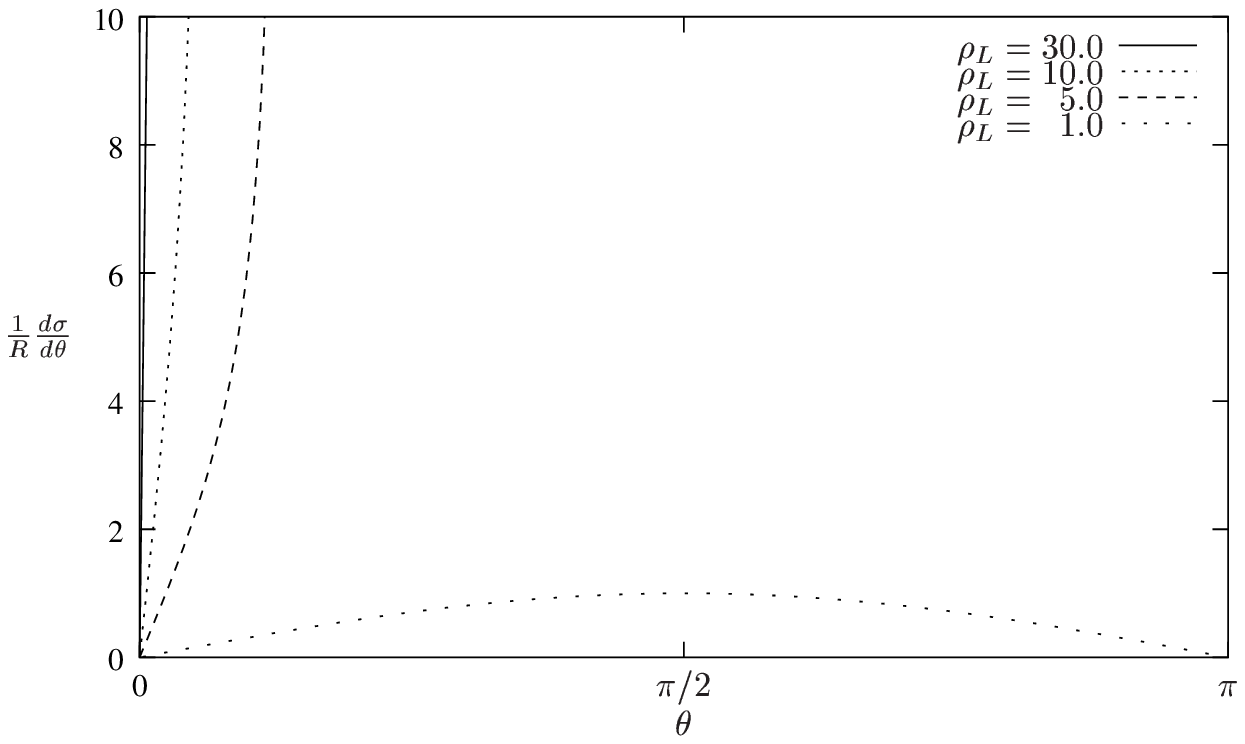,bb=119 453 476 667,angle=0,width=\linewidth,clip=}
 \caption[Classical cross section for the scattering by a solenoidal magnetic field for several values of
          $\rho_L = pR/2\beta \ge  1$. ]
         {Classical cross section for the scattering by a solenoidal magnetic field for several values of
          $\rho_L = pR/2\beta \ge 1$.}
 \label{figure:classical_CS_more}
\end{figure}
\end{center}

Figure~\ref{figure:classical_CS_more} shows the behavior of the
classical DCS for some values of the parameter $\rho_L$ when it is
greater than unity. As can be observed, the DCS behaves each time more
singular and the maximum scattering angle tends to zero as the
parameter $\rho_L$ increases.

%%%%%%%%%%%%%%%
%%%%%%%%%%%%%%%

%% Subsection: The asymmetry function
\subsection{The asymmetry function}

As stated before the probability to find scattered particles in the
upper positive half-plane is in general different from the
one to find scattered particles in the lower
negative half-plane.
To show this explicitly let us introduce the asymmetry of the cross
section as a function of the parameter $\rho_L$,
$$
A(\rho_L) = \frac{\sigma_{+}(\rho_L) - \sigma_{-}(\rho_L)}
                      {\sigma(\rho_L)},
$$ 
with $\sigma(\rho_L) = \sigma_{+}(\rho_L) + \sigma_{-}(\rho_L)$.
$\sigma_{\pm}$ correspond to the total cross
section for $0 \leq \theta < \pi$ and $\pi \leq \theta < 2\pi$,
respectively.
For all values of $\rho_L$, $\sigma(\rho_L) = 2R$.
As it is evident from equation~(\ref{eq:secc_dif_clasica}),
for $\rho_L \geq 1$ the cross section 
is completely asymmetric. In this case
\begin{eqnarray*}
\sigma_{-}(\rho_L \geq 1) &=& 0, \\
\sigma_{+}(\rho_L \geq 1) &=& \int_{0}^{\theta_{max}}{\frac{d\sigma_{\rho_L}(\theta)}{d\theta}\, d\theta} 
                           =  2 R,
\end{eqnarray*}
and
$$
A(\rho_L \geq 1) = 1;
$$
whereas for  $\rho_L < 1$, 
\begin{eqnarray*}
  \sigma_{+}(\rho_L < 1) &=& \int_{0}^{\pi}{\frac{d\sigma_{\rho_L}(\theta)}{d\theta}\, d\theta} = R\,(\rho_L + 1), \\  
  \sigma_{-}(\rho_L < 1) &=& \int_{\pi}^{2\pi}{\frac{d\sigma_{\rho_L}(\theta)}{d\theta}\, d\theta} = R\,(1 - \rho_L). 
\end{eqnarray*}
In this form, 
\begin{equation}
A(\rho_L<1) = \rho_L.
\label{eq:asimetria_clasica_rho<1}
\end{equation}

It follows from equation~(\ref{eq:asimetria_clasica_rho<1}) that
only in the limit 
$\rho_L \rightarrow 0$ the cross section is completely
symmetric, $A(\rho_L\rightarrow 0) \rightarrow 0$. In such a
case, for fixed solenoid radius, the Larmor radius of the scattered particles is smaller than the
solenoid one and backscattering is greatly favored. 

\begin{center}
\begin{figure}[ht]
\epsfig{file=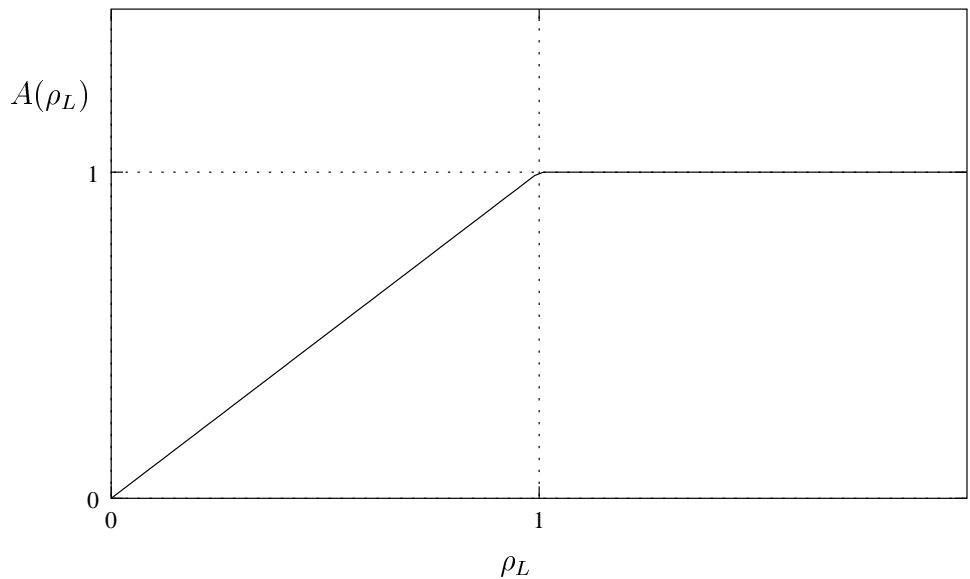,bb=127 449 491 667,angle=0,clip=}
 \caption[Asymmetry function of the classical cross section of the scattering by a solenoidal magnetic field.]
         {Asymmetry function of the classical cross section of the scattering by a solenoidal magnetic field.}
 \label{figure:assimetry}
\end{figure}
\end{center}

Figure~\ref{figure:assimetry} depicts the general behavior of $A(\rho_L)$.
%%%%%%%%%%%%%%%
%%%%%%%%%%%%%%%

%%% Section: Quantum mechanical results %%%  
\section{Quantum mechanical results}
\label{sec:Quantum_Result}  

Both the relativistic and non-relativistic quantum mechanical
scenarios for the scattering of charged particles by a solenoidal
magnetic field are presented below. The non-relativistic case can be
exactly solved in the zero radius limit. We briefly recall the most
important results obtained years ago by Aharonov and Bohm~\cite{AB}
and Landau and Lifshitz~\cite{LL} in the non-relativistic regime.
The more general case of finite non-zero radius of the
solenoid is discussed in the relativistic regime section using first
order perturbation theory.

%% Subsection: Non relativistic regime%%%  
\subsection{Non relativistic regime}  

A landmark result for the non relativistic 
scattering of electrons by solenoidal magnetic fields
was presented by Aharonov and Bohm~\cite{AB}. They
obtained the exact solution for the scattering problem in the zero
radius limit of the solenoid for a constant magnetic flux $\Phi$,
\begin{equation}  
\left.{\frac{d\sigma}{d\theta}}\right|_{AB}=  
         \hbar \, \frac{\sin^2{\left(e\Phi/2\hbar c\right)}}  
         {2 \pi p \sin^2{(\theta/2)}}.
\label{eq:secc_dif_AB}  
\end{equation}
This result is manifestly symmetric in the scattering angle under the
$\theta \rightarrow 2\pi - \theta$ transformation.
  
Landau and Lifshitz~\cite{LL} studied the same
scattering problem in the eikonal
approximation. Including only the contribution of the vector potential
from the exterior of the solenoid, they obtained precisely the same result as
Aharonov and Bohm.  These authors also studied the case of
small scattering angles for a small magnetic
flux, $e\Phi/2\hbar c \ll 1$, where perturbation theory is applicable,
the resulting cross section is again symmetric in the
scattering angle with respect to the incident direction of the particles ($\theta = 0$):
\begin{equation}  
\left.{\frac{d\sigma}{d\theta}}\right|_{LL; \theta \ll 1} =  
       \hbar \, \frac{\left(e\Phi/\hbar c\right)^2}{2\pi p} \, \frac{1}{\theta^2}.  
\label{eq:secc_dif_LL_singular}  
\end{equation}  
They mention that the singular behavior of the total cross section for
$\theta$ going to zero is specifically a quantum effect, without any
further comment.

In any case the limit $R=0$, clearly separates the classical
regime (in which $d\sigma/d\theta$ cancels) with respect to the
quantum regime (equations~(\ref{eq:secc_dif_AB})
and~(\ref{eq:secc_dif_LL_singular})). Consequently any further
comparison of the classical and quantum solutions requires to consider
the finite radius situation.
%

%% Subsection: Relativistic regime %%%
\subsection{Relativistic regime}
\label{sec:Relativistic_regime}

The relativistic quantum mechanical problem of the scattering of electrons by the
magnetic field of a long solenoid of radius $R$ with axial axis in the
$\bi{x_3}$ direction is considered here.  The magnetic flux
$\Phi = \pi R^2 B_0$ will be kept constant.

Once the gauge has been fixed, the magnetic vector potential of the
solenoid is
\begin{equation}    
\slash{A}=A_{\mu}\gamma^{\mu}=\frac{\Phi}{2\pi}\epsilon_{ij3}x_i\gamma^j  
\cases{\frac{1}{R^2} & for $r<R$ \cr  
            \frac{1}{x^2_1+x^2_2} & for $r>R$,}  
\label{eq:solenoidal_potential}
\end{equation}
with scalar potential $A^0=0$.  
$\epsilon_{ijk}$ is the Levi-Civita symbol in three indexes. 

The first order matrix element in $e$, $S_{fi}^{(1)}$, that has to be computed is
$$
S_{fi}^{(1)} =   
 \delta_{fi} -   
  ie\int{\bar{\psi}_f(y)\slash{A}(y)\psi_i(y) d^4y},  
$$
with $\psi_i(y)$ and $\psi_f(y)$ free particle incident and final
asymptotic states. A detailed calculation of the first order Born
approximation for the DCS yields~\cite{Murguia_2003}
\begin{equation}   
\frac{d\sigma}{d\theta} =   
       \hbar \left({\frac{e\Phi}{Rc}}\right)^2   
       \frac{{\left| J_1(2\frac{p}{\hbar}R  
                     {\left|\sin{(\theta/2)}\right|})  
              \right|}^2}  
            {8\pi p^3 \sin^4{(\theta/2)}}, 
\label{eq:dsigma_MM}   
\end{equation}
where $J_1$ are the first order Bessel functions of first kind.
The previous result has the same form whether or not the final polarization of the
beam is actually measured. As can be observed, the lowest perturbative
order in $\alpha = e^2/\hbar c$ of the relativistic quantum mechanical
cross section is also symmetric in $\theta$.

An analysis~\cite{Murguia_2003} of the classical Planck's
limit (taking $\hbar \rightarrow 0$) of
equations~(\ref{eq:secc_dif_AB}) and~(\ref{eq:dsigma_MM}), with fixed
$e, p, R, \Phi$ and $\theta$, yields a consistent zero limit,
contrary to the asymmetric finite value of the classical DCS in
equation~(\ref{eq:secc_dif_clasica}).
From equation~(\ref{eq:dsigma_MM}) we have
\begin{equation}  
\lim_{\hbar \rightarrow 0}\frac{d\sigma}{d\theta}  
  =\lim_{\hbar \rightarrow 0} \hbar^2   
   \left({\frac{e\Phi}{2\pi c}}\right)^2  
   \frac{\cos^2{\left({2\frac{p}{\hbar}R  
                      {\left|\sin{(\theta/2)}\right|}   
                      - 3\pi/4 }\right)}}  
        {2 R^3 p^4 \left|{\sin^5{(\theta/2)}}\right|}  
  =0.
\label{eq:classicla_limit-dsigma_MM}  
\end{equation}   
Notice that the same limit is recovered by taking 
$p R \rightarrow \infty$ with $\hbar$ and $R$ fixed~\cite{Skarzhinsky_1997}.

Equation~(\ref{eq:classicla_limit-dsigma_MM}) establishes that, to
first order in $\alpha$, there is not a classical correspondence of
the phenomena of scattering of charged fermions by the magnetic field
of a solenoid with finite non-zero radius. It is instructive to
compare this with the Coulomb scattering, in which both the
relativistic and non-relativistic quantum results reproduce the well
known Rutherford classical cross section in the classical limit $\hbar
\rightarrow 0$. Notice that the classical-quantum correspondence for
the Coulomb field takes place in a perturbative regime to first order
in $\alpha$.

\begin{center}
\begin{figure}[ht]
\epsfig{file=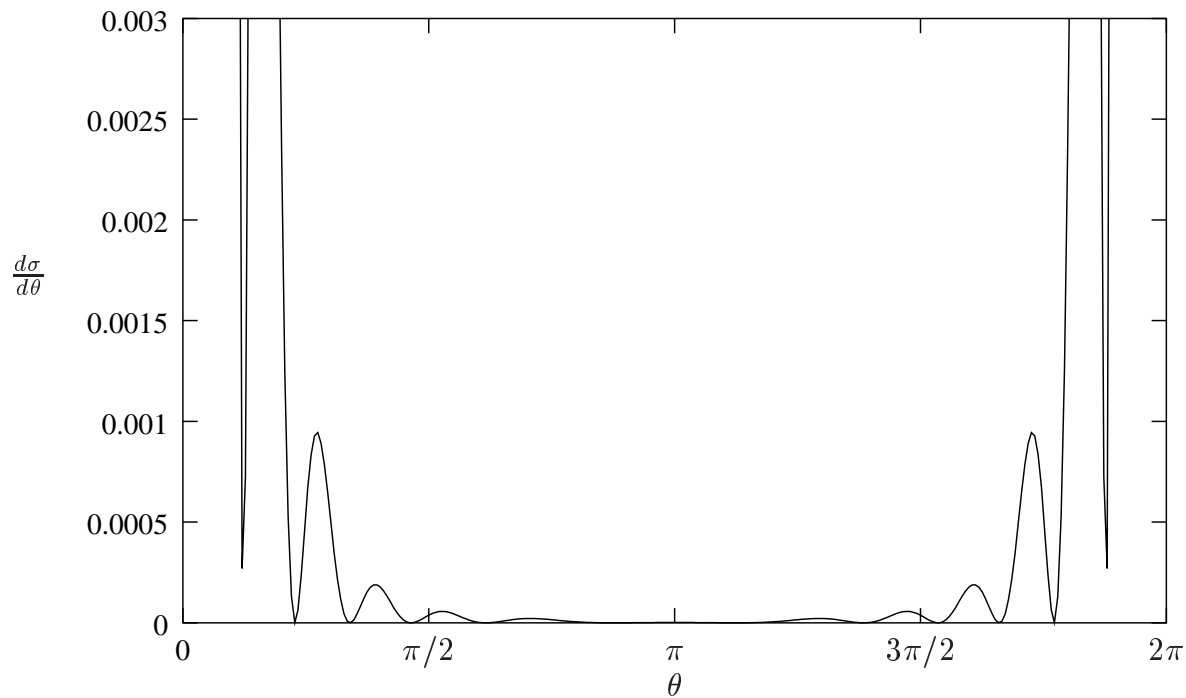,bb=123 453 483 667,angle=0,width=\linewidth,clip=}
 \caption[Typical behavior of the relativistic quantum cross section of 
          the scattering by a solenoidal magnetic field.]
         {Typical behavior of the relativistic quantum cross section of 
          the scattering by a solenoidal magnetic field.}
 \label{figure:quantum_cross_section}
\end{figure}
\end{center}

In this quantum relativistic regime, the asymmetry function is also
equal to zero, because $d\sigma(2\pi - \theta) = d\sigma(\theta)$ and
therefore $\sigma_{-} = \sigma_{+}$.
Figure~\ref{figure:quantum_cross_section} shows the symmetric
behavior of the relativistic quantum cross section of
equation~(\ref{eq:dsigma_MM}). Notice that the scattering is directed
mainly in the forward direction.
%%%%%%%%%%%%%%%
%%%%%%%%%%%%%%%

The essentially different behavior between the classical and quantum
DCS becomes evident from the symmetric behavior of the quantum result,
equation~(\ref{eq:dsigma_MM}), as compared to the asymmetric
structure of the classical one, equation~(\ref{eq:secc_dif_clasica}). 
Furthermore, notice that the total cross section in both quantum
regimes is infinite, in contrast to the finite value of $2R$ obtained
for the classical case.

In order to throw further understanding of these facts, we notice that
the classical result depends on two length-dimensional parameters: the
solenoid radius $R$ and the Larmor radius $r_L$. In the quantum regime
we can add two more length-dimensional parameters, the de Broglie
length $\lambdabar = \hbar/p$ and the magnetic length $\ell_B =
\sqrt{\hbar c/eB} = R \sqrt{\hbar c \pi/e \Phi}$. Not all these
parameters are independent, in fact $r_L = \ell_B^2/\lambdabar$. The
quantum DCS in equation~(\ref{eq:dsigma_MM}) can be expressed in terms
of three of these parameters. {\it e.g.} $R, r_L$ and
$\lambdabar$. Unlike the Rutherford formula, the cross section in
equation~(\ref{eq:dsigma_MM}) can not be written purely in terms of
classical length parameters.

As it has been noticed throughout this paper, $\rho_L = r_L/R$ results
to be the significant parameter in the classical regime. In the
quantum case, we can identify two action-dimension parameters: $pR$
and $e\Phi/c$.  A comparison of these parameters respect to $\hbar$
helps to have a better understanding of the structure of the quantum
DCS.  In terms of the dimensionless parameters, $s_p = pR/\hbar$
and $s_\Phi = e\Phi/\hbar c$, the DCS of Aharonov and Bohm (AB-DCS),
equation~(\ref{eq:secc_dif_AB}), looks like
$$
\left.{\frac{d\sigma}{d\theta}}\right|_{AB}=  
         R \, \frac{\sin^2{(s_\Phi/2)}}  
                {2 \pi s_p \sin^2{(\theta/2)}},
$$ 
while the perturbative relativistic DCS of
equation~(\ref{eq:dsigma_MM}) can be expressed as
\begin{equation}
\frac{d\sigma}{d\theta} =   
       \frac{R}{8\pi} \frac{s_\Phi^2}{s_p^3}
       \left|{ \frac{J_1(2 s_p {\left|\sin{(\theta/2)}\right|})}
                    {\sin^2{(\theta/2)}} 
      }\right|^2.
\label{eq:DCS_MM_s_Phi-s_p}
\end{equation}
The former equation results from a perturbative expansion in $s_\Phi$;
it is therefore obtained for arbitrary values of $s_p$, but small
$s_\Phi$.  On the other hand, the range of validity of the AB-DCS is
arbitrary $s_\Phi$ but small $s_p$ (in fact, it was obtained in the
special case of $s_p=0$).

In terms of the same parameters $s_p$ and $s_\Phi$, $\ell_B = R
\sqrt{\pi/s_\Phi}$, $\lambdabar = R/s_p$ and $\rho_L = \pi
s_p/s_\Phi$, and the classical DCS can be expressed as a function of
the ratio $s_p/s_\Phi$ and $\theta$ only.
 
Thus the results in~(\ref{eq:secc_dif_AB}) and~(\ref{eq:dsigma_MM}) do
not reduce to the classical result in the $\hbar \rightarrow 0$
limit. One may wonder if there is still a situation in which it is
possible to establish a correspondence between the quantum and
classical results. It becomes relevant to observe, that according to
Berry and Mount~\cite{Berry_1972} and
Gutzwiller~\cite{book:Gutzwiller}, the implementation of the classical
limit requires to look at the situation in which the action quantities
that appear in the corresponding classical problem are considered very
large as compared to $\hbar$~\cite{Note_References}.
In the problem at hand we can select $s_p$ and $s_\Phi$ as the
relevant parameters.
Consequently, the classical limit of the relativistic DCS vanishes,
because for $s_p$ arbitrary large,
equation~(\ref{eq:DCS_MM_s_Phi-s_p}) behaves like $s_\Phi^2/s_p^4$. A
similar situation happens for the non-relativistic case.

%%% Section: Quantum vs Classical results%%   
\section{Quantum vs Classical results}  
\label{sec:Analysis}  

Here we extend the discussion of the $\hbar$ dependence of the
relativistic DCS to all orders in $\beta = e\Phi/2\pi c$ and $\alpha=
e^2/\hbar c$, a consideration that becomes crucial for the inspection
of the classical limit of the scattering process.

The consistency between the non-relativistic and the relativistic
quantum results has been well established before~\cite{Murguia_2003, deAlbuquerque,
Boz, Shikakhwa} in the $R \rightarrow 0$ limit. However, we want to
stress that there is still an unsolved problem with the classical
Planck's limit that shows itself clearly in the lack of asymmetry of
the quantum calculations.

\begin{figure}[ht]
\begin{center}
\epsfig{file=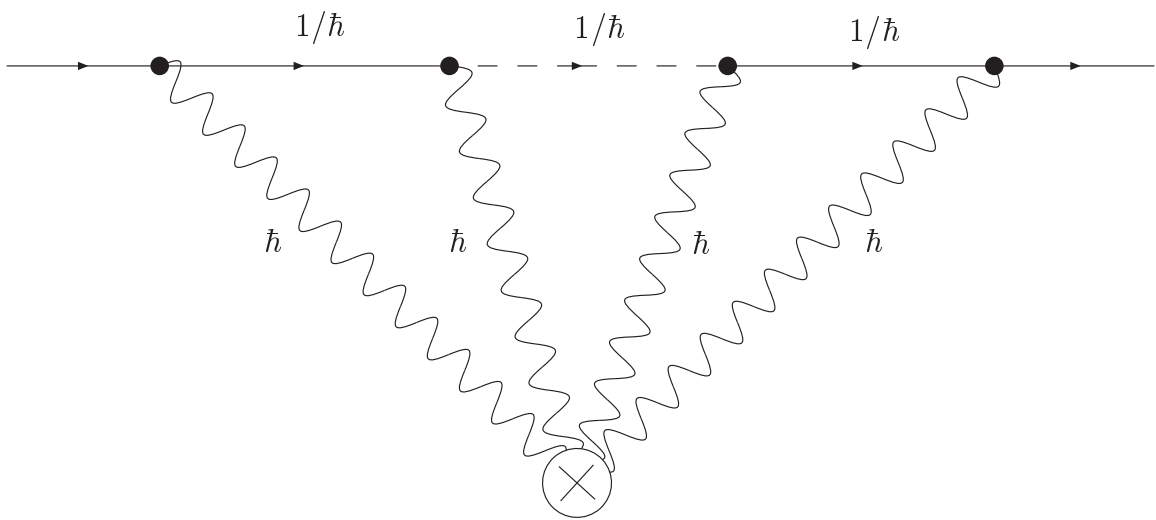,bb=71 490 580 720,angle=0,width=\linewidth,clip=}
\end{center}
 \caption[Feynman diagram and $\hbar$ power counting for an arbitrary
          order in $\beta = e\Phi/2\pi c$ of the scattering matrix for
          a solenoidal magnetic field.  The wiggled lines
          represent the interaction with the
          external magnetic field while the straight lines represent the
          free-fermion propagators.]
         {Feynman diagram and $\hbar$ power counting for an arbitrary
          order in $\beta = e\Phi/2\pi c$ of the scattering matrix for
          a solenoidal magnetic field. The wiggled lines
          represent the interaction with the
          external magnetic field while the straight lines represent the
          free-fermion propagators.}
 \label{figure:Feynman_diagram_all_orders}
\end{figure}

Higher order processes can be calculated using the Feynman rules for
the electron-solenoid scattering that are presented in the
appendix. Here, we are interested in counting the $\hbar$ power
contributions to higher order diagrams as the ones depicted in
figure~\ref{figure:Feynman_diagram_all_orders}.
Assuming free particle asymptotic states, we notice that any extra
insertion of the external magnetic field ({\it i.e.} in $\beta =
e\Phi/2\pi c$) to the scattering matrix contributes with a magnetic
interaction and a free-fermion-propagator (see
figure~\ref{figure:Feynman_diagram_all_orders}).  Each magnetic
interaction contributes with a $n=+1$ power of $\hbar$, while for each
free fermion propagator there is a $n=-1$ power in addition to the
global $\hbar$ factor of the scattering matrix in momentum space (see
the appendix). 
This means that higher orders in $\beta$ do not modify the leading
$\hbar$ power contribution to the scattering matrix.

%%%%%%%%%%%%%%%
%%%%%%%%%%%%%%%

We recall that as is well known, usual radiative corrections (higher
powers in $\alpha = e^2/\hbar c$) will in general contribute with
positive $\hbar$ powers to the matrix elements, hence they are not
expected to be relevant in the classical limit.

Consequently, for all orders in both $\alpha$ and $\beta$ in the
perturbative expansion, the classical Planck's limit of this process
is proportional to $\hbar$ and the classical result can not be
recovered with a perturbative calculation in $\beta$ and $\alpha$.  As
discussed in the last section, the same zero classical limit is found
analyzing the DCS behavior in terms of the $s_\Phi$ and $s_p$
variables.

In the quantum regime the perturbative cross section has a symmetric
behavior with respect to the scattering angle when it changes form
$\theta$ to $2\pi - \theta$, whereas in the classical scenario, the
symmetry in the scattering angle only occurs in the limit $pR
\rightarrow 0$ ($\rho_L \rightarrow 0$).  Higher orders in $\beta$ are
expected to give rise to up-down asymmetries with respect to the
scattering angle in the quantum DCS for the finite solenoid radius
case, but notice that as we stated before, the renormalized series in
$\beta$ of the scattering matrix leads to a null classical
limit. Therefore these asymmetries in the perturbative quantum
calculations clearly are not the classical ones.

In this sense, the renormalized perturbative calculation of scattering
of Dirac particles by a solenoidal magnetic field has to be understood
as a pure quantum one.
Only an exact non-perturbative calculation at finite radius $R$ could
possible produce a consistent result from which the classical limit
could be possible recovered.

%%% Conclusions %%%
\section{Conclusions}  
\label{sec:Conclusions}

In this paper we present an analysis for the scattering of charged
particles by a solenoidal magnetic field of finite radius $R$ and
constant magnetic flux $\Phi$ in both, classical and quantum
relativistic and non-relativistic regimes. We focus on the classical
Planck's limit, taking $\hbar \rightarrow 0$, of the quantum results
in the framework of a perturbation theory in $\alpha = e^2/\hbar c$
and $\beta = e\Phi/2\pi c$.

In order to have a clear reference to compare with, and as we did not
find it reported in the literature, we explicitly calculated the
differential cross section (DCS) for the classical scenario. There, we
found a general asymmetric behavior of the DCS with respect to the
scattering angle $\theta$. This result differs with respect to the
symmetric DCS obtained in the quantum regime.
Also notice that in the quantum regime the total cross section of
the scattering problem becomes infinite (for both cases, $R \neq 0$
and $R=0$), while the classical total cross section is finite and
equal to $2R$.

As we have showed, the quantum results for the DCS are proportional to
$\hbar$.  In this paper we studied the contribution to the power
counting of $\hbar$ of the perturbative expansion to all orders in
both, $\alpha$ and $\beta$, showing that in the classical limit,
in the Planck's sense, the leading term will be proportional
to $\hbar$, so it vanishes.
This means that the perturbative evaluation of the scattering matrix
can not converge to the classical solution in the $\hbar \rightarrow
0$ limit, this one understood as the limit in which all the classical
action variables are very large as compared to the Planck's constant
$\hbar$.
Our conclusion is that only an exact non-perturbative calculation for
a finite solenoid radius, offers the possibility to obtain a
consistent result for the DCS, from which the classical limit could be
possible recovered.

%%% Acknowledgments %%%  
\ack 

This work was partially supported by 
CONACyT grants G32723E and 42026-F, 
DGAPA-UNAM grants IN120602 and IN118600 and 
DGEP-UNAM.

%%% Appendix %%%  
\appendix
\section*{Appendix. Feynman rules for the solenoidal case}  
\label{sec:Appendix:Feynman_Rules}  

The general Feynman rules for the scattering amplitude of electrons by
a solenoidal magnetic field of finite non-zero radius $R$ are presented here.

Notice first that in standard perturbation theory, using free particle
asymptotic states, the lowest order contribution in $e$ to the scattering
matrix is given by the Fourier transformation of the four-vector potential $A^\mu(x)$:
$$
S_{fi}^{(1)} =   
 \delta_{fi} - i \int{\bar{\psi}_f(x) \frac{e \slash{A}(x)}{\hbar c} \psi_i(x) d^4x},  
$$
with
$$
\psi(x) =  \sqrt{\frac{mc^2}{EV}}u({\bi p},{\bi s})e^{-i p \cdot x /\hbar}.  
$$

For the solenoidal potential given by equation~(\ref{eq:solenoidal_potential}),
the first order matrix element results in
$$
S_{fi}^{(1)} = \sqrt{\frac{m_ic^2}{E_i V}} \sqrt{\frac{m_fc^2}{E_f V}}
               \left(\frac{e\Phi}{\hbar c}\right) \bar{u}_f 
               \left[ - 2 i \frac{\hbar^2}{R} J_1(q_\perp R/\hbar) \epsilon_{ij3} \frac{q_i}{q_\perp^3} \gamma^j\right]
               u_i
	       (2\pi)^2 \delta^2(q_\parallel/\hbar),
$$ 
where $q = p_f - p_i$ is the transfered momentum. We have defined
$a^\mu = a^\mu_\perp + a^\mu_\parallel$, with $ a^\mu_\perp =
(0,a_1,a_2,0)$ and $a^\mu_\parallel = (a_0,0,0,a_3)$, where the
subscripts $\perp$ and $\parallel$ refer to the components of the
four-vector that are perpendicular or parallel to the direction of the
solenoidal magnetic field.

\begin{figure}[ht]
\begin{center}
\epsfig{file=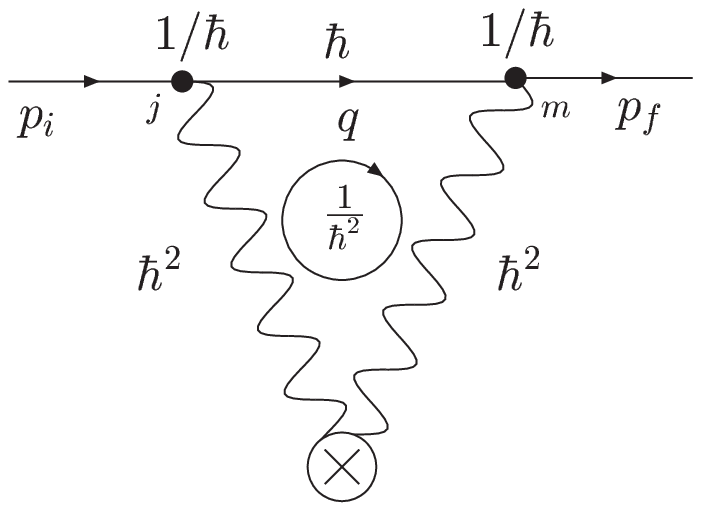,bb=187 569 408 721,angle=0,clip=}
\end{center}
 \caption[Second order Feynman diagram of the scattering matrix for
          a solenoidal magnetic field.]
         {Second order Feynman diagram of the scattering matrix for
          a solenoidal magnetic field}
 \label{figure:Feynman_diagram_Second_Order}
\end{figure}

$S_{fi}^{(1)}$ has the structure of the product: {\it final
state spinor} $\times$ {\it magnetic-interaction vertex} $\times$
{\it magnetic propagator} $\times$ {\it initial state spinor}
$\times$ {\it energy-momentum conservation}.
Precisely this structure determines the Feynman rules of the invariant
amplitude ${\cal M}$ of the problem:

\begin{itemize}

  \item Each external incoming (anti)fermionic line contributes with a
  factor $u({\bi p},{\bi s})$ ($v({\bi p},{\bi s})$).

  \item Each external out-coming (anti)fermionic line contributes with a
  factor $\bar{u}({\bi p},{\bi s})$ ($\bar{v}({\bi p},{\bi s})$).

  \item For each magnetic-fermionic vertex a dimensionless factor appears:
  $$
    \frac{e\Phi}{\hbar c}.
  $$

  \item Each magnetic propagator contributes to ${\cal M}$ with
  $$
    - 2 i \frac{\hbar^2}{R} J_1(q_\perp R/\hbar) \epsilon_{ij3} \frac{q_i}{q_\perp^3} \gamma^j,
  $$ 
  which has the dimensions of length.
  This propagator will contribute with a factor of $\hbar^2$ to the
  amplitude. Notice that this is a general result for any
  electromagnetic interaction in a bi-dimensional problem.

  \item For each internal fermionic line, the free fermion propagator will appear:
  $$
    - i S_F(q) = - i\, \frac{\hbar}{\slash{q} - mc +i\epsilon}.
  $$ 
  Additionally there is 
  an integral in $d^4q/\hbar^4$ and due to the
  cylindrical symmetry two of this variables are integrated out
  immediately using the momentum-energy conservation (see the next
  rule). Then, $S_F(q)$ will finally contribute with a $1/\hbar$
  factor to the amplitude. This situation is equivalent to consider an
  internal loop in the $q_\perp$-space (see
  figure~\ref{figure:Feynman_diagram_Second_Order}). 

  \item For each magnetic-fermionic vertex, both the energy and the
  momentum along the direction of the magnetic field, are
  conserved. So, the additional term 
  $$
    (2\pi)^2 \delta^2(q_\parallel/\hbar),
  $$
  arises in the amplitude.

\end{itemize}

Additional terms have to be considered along with the phase space
factors to construct the differential cross section. The normalization
factors of the external fermionic lines have to be included in these
extra-terms.

As an example, let us apply the Feynman rules just listed before to
write down the second order matrix element of the scattering process,
${\cal M}^{(2)}$; 
figure~\ref{figure:Feynman_diagram_Second_Order} depicts the
Feynman diagram that has to be evaluated.
There, the $\hbar$ power contribution to ${\cal M}^{(2)}$ is shown
explicitly for each part of the diagram.

%%%%%%%%%%%%%%%
%%%%%%%%%%%%%%%

According to the Feynman rules already established:
$$
{\cal M}^{(2)} = \cases{ \frac{1}{(2\pi)^4} \int \frac{d^4q}{\hbar^4} \left\{  \right.
                        \bar{u}({\bi p_f},{\bi s_f})    & Out-coming external fermion \cr
         \times \left( \frac{e\Phi}{\hbar c} \right)^2  & Vertexes $m$ and $j$ \cr
         \times \left[- 2 i \frac{\hbar^2}{R} J_1(|p_f-q|_\perp R/\hbar) \epsilon_{lm3} \frac{(p_f-q)_l}{|p_f-q|_\perp^3} \gamma^m\right]%
                                                        & Magnetic propagator at $m$ \cr
         \times \left[- i\, \frac{\hbar}{\slash{q} - mc +i\epsilon}\right]%
                                                        & Free fermion propagator \cr
         \times \left[- 2 i \frac{\hbar^2}{R} J_1(|q-p_i|_\perp R/\hbar) \epsilon_{ij3} \frac{(q-p_i)_i}{|q-p_i|_\perp^3} \gamma^j\right]%
                                                        & Magnetic propagator at $j$ \cr
         \times          u({\bi p_i},{\bi s_i})         & Incoming external fermion  \cr
         \times \left[ (2\pi)^2 \delta^2((p_f-q)_\parallel/\hbar) \right] \left[ (2\pi)^2 \delta^2((q-p_i)_\parallel/\hbar) \right]%
                                                                      \left. \right\}%
                                                        & Energy-momentum conservation, }
$$
which with the aid of a pair of the Dirac-delta functions reduces to
\begin{eqnarray*}
{\cal M}^{(2)} = & &  (-i)^3 \left( \frac{2 e\Phi}{\hbar c R} \right)^2 (2\pi)^2 \delta^2[(p_f-p_i)_\parallel/\hbar]%
                             \, \epsilon_{lm3} \epsilon_{ij3} \\
                 & &  \times \frac{1}{(2\pi)^2} \int \frac{d^2q_\perp}{\hbar^2} \, %
                             \bar{u}_f \, \gamma^m \frac{\hbar}{\slash{p}_\parallel + \slash{q}_\perp - mc +i\epsilon} \gamma^j \, u_i \\
		 & &  \times \hbar^2 \, \frac{(p_f-q)_l}{|p_f-q|_\perp^3} \, J_1(|p_f-q|_\perp R/\hbar) \,%
                             \hbar^2 \, \frac{(q-p_i)_i}{|q-p_i|_\perp^3} \, J_1(|q-p_i|_\perp R/\hbar).
\end{eqnarray*}
In terms of the dimensionless action-related variables $s_p=pR/\hbar$ and
$s_\Phi=e\Phi/\hbar c$, this last expression reads
\begin{eqnarray*}
{\cal M}^{(2)} = & &  (-i)^3 R \, (2 s_\Phi)^2 (2\pi)^2 \delta^2[(p_f-p_i)_\parallel/\hbar]%
                             \, \epsilon_{lm3} \epsilon_{ij3} \\
                 & &  \times \frac{1}{(2\pi)^2} \int d^2s_{q_\perp} \, %
                             \bar{u}_f \, \gamma^m \frac{1}{\slash{s}_{p_\parallel} + \slash{s}_{q_\perp} - mRc +i\epsilon} \gamma^j \, u_i \\
		 & &  \times \frac{(s_{p_f}-s_q)_l}{|s_{p_f}-s_q|_\perp^3} \, \frac{(s_q-s_{p_i})_i}{|s_q-s_{p_i}|_\perp^3} \, %
                             J_1(|s_{p_f}-s_q|_\perp) \, J_1(|s_q-s_{p_i}|_\perp).
\end{eqnarray*}

%%% References %%%

\end{document}